\title{Computing the diameter polynomially faster than APSP}

\author{
\em Raphael Yuster
\thanks{Department of Mathematics, University of Haifa, Haifa
31905, Israel. E--mail: raphy@math.haifa.ac.il}
}

\date{}

\documentclass [letterpaper,11pt]{article}
\usepackage{amsfonts}

\newtheorem{theorem}{Theorem}[section]

\newtheorem{lemma}[theorem]{Lemma}
\newtheorem{definition}[theorem]{Definition}

\newcommand{\qed}{\hspace*{\fill} \rule{7pt}{7pt}}
\newcommand{\Proof}{\noindent{\bf Proof.}\ \ }

\newcommand{\w}{\omega}
\newcommand{\Ot}{\tilde{O}} 

\begin{document}
\maketitle

\begin{abstract}

We present a new randomized algorithm for computing the diameter of a weighted directed graph.
The algorithm runs in $\Ot(M^{\w/(\w+1)}n^{(\w^2+3)/(\w+1)})$ time, where $\w < 2.376$ is the exponent
of fast matrix multiplication, $n$ is the number of vertices of the graph, and
the edge weights are integers in $\{-M,\ldots,0,\ldots,M\}$.
For bounded integer weights the running time is $O(n^{2.561})$ and if $\w=2+o(1)$
it is $\Ot(n^{7/3})$.
This is the first algorithm that computes the diameter of an integer weighted directed graph polynomially faster than
any known All-Pairs Shortest Paths (APSP) algorithm. For bounded integer weights, the fastest algorithm for APSP runs in $O(n^{2.575})$ time for the present value of $\w$ and runs in $\Ot(n^{2.5})$ time if $\w=2+o(1)$.

For directed graphs with {\em positive} integer weights in $\{1,\ldots,M\}$ we obtain a deterministic algorithm that
computes the diameter in $\Ot(Mn^\w)$ time. This extends a simple $\Ot(n^\w)$ algorithm for computing the diameter
of an {\em unweighted} directed graph to the positive integer weighted setting and is the first algorithm in this setting whose time complexity matches that of the fastest known Diameter algorithm for {\em undirected} graphs. 

The diameter algorithms are consequences of a more general result.
We construct algorithms that for any given integer $d$,
report all ordered pairs of vertices having distance {\em at most} $d$.
The diameter can therefore be computed using binary search for the smallest $d$ for which all pairs are reported.

%{\bf Key words.} fast matrix sparsification.

%{\bf AMS subject classifications.} 68W30

\end{abstract}

\section{Introduction}\label{s:intro}

Computing the diameter and, more generally, computing distances, are among the most fundamental algorithmic
graph problems. In the {\em Diameter} problem we are given a graph
and are required to find the largest distance between two vertices of the graph.
The more general {\em All-Pairs Shortest Paths} (APSP) problem asks to
find distances and shortest paths between all pairs of vertices of the graph.
Clearly, any algorithm for APSP can be used as an algorithm for Diameter.
The converse, however, is not necessarily true.

Unfortunately, at present, we do not know of any algorithm for {\em general} weighted graphs that solves Diameter asymptotically faster than APSP
and the existence of such an algorithm is an open problem (see, e.g., \cite{ACIM99,Ch06,Ch87}).

In the case of graphs with arbitrary real edge weights, no truly sub-cubic algorithm for APSP or for
Diameter is known. The presently fastest algorithm for both is an algorithm of Chan \cite{Ch07} that runs
in $O(n^3\log^3\log n/\log^2n)$ time, where $n$ in the number of vertices of the graph. For sufficiently sparse graphs, the APSP algorithm of Johnson \cite{Jo77} performs
better as it runs in $O(mn+n^2 \log n)$ time where $m$ is the number of edges.

When the edge weights are integers, fast matrix multiplication techniques are useful.
For undirected graphs with integer edge weights in $\{1,\ldots,M\}$ an APSP (and thereby a Diameter) algorithm of Shoshan and Zwick
\cite{ShZw99} runs in $\Ot(Mn^w)$\footnote{Throughout this paper, $\Ot(f(n))$ stands for $f(n)n^{o(1)}$.} time,
where $\w < 2.376$ is the exponent of matrix multiplication. This algorithm generalizes earlier $\Ot(n^\w)$
algorithms of Seidel and of Galil and Margalit for the unweighted case \cite{Se95,GaMa97}.

The situation becomes more involved in directed graphs, as negative edge weights may be allowed.
In the presence of weights in $\{-M,\ldots,0,\ldots,M\}$,
the fastest APSP algorithm (and presently fastest Diameter algorithm) is by Zwick \cite{Zw02}.
It runs in $\Ot(M^{1/(4-\w)}n^{2+1/(4-\w)})$ time, and an additional
small speedup is obtained if fast rectangular matrix multiplication is used.
For bounded $M$, this speedup results in a running time of $O(n^{2.575})$.
Interestingly, Zwick's algorithm is also the fastest known algorithm for {\em unweighted} APSP in
directed graphs. The directed unweighted setting is also the only known setting where Diameter has an
algorithm that is presently faster than APSP. It was observed by Zwick (private communication) and
also by Bj\"orklund (private communication) that the Diameter of a directed unweighted graph can be computed
in $\Ot(n^\w)$ time using a repeated squaring argument of the adjacency matrix, combined with binary search.
This approach, however, does not directly extend to the weighted setting even when the weights
are positive integers in $\{1,\ldots,M\}$.

The main results of this paper present the first algorithm(s) for Diameter that are polynomially faster than APSP
in integer weighted directed graphs. Our first algorithm computes the diameter of a directed graph with integer edge weights
in $\{-M,\ldots,0,\ldots,M\}$ polynomially faster than the aforementioned APSP algorithm of Zwick.
Our second algorithm computes the diameter of a directed graph with integer edge weights
in $\{1,\ldots,M\}$ even faster. The time complexity we obtain in this case matches the time complexity of the aforementioned APSP algorithm of Shoshan and Zwick that applies to the {\em undirected} setting.

In fact, both algorithms are, respectively, consequences of two other algorithms that solve a related problem that can be viewed as more general than Diameter.

The problem we are considering is {\em Threshold APSP} where, in addition to the input graph we receive a threshold
value $d$. The goal is to report all ordered pairs of vertices having distance {\em at most} $d$.
Any algorithm for Threshold APSP can be converted into an algorithm for Diameter by applying
Threshold APSP a logarithmic number of times. Indeed, since our edge weights are integers in $\{-M,\ldots,0,\ldots, M\}$
the diameter is either $\infty$ or is an integer in $\{0,\ldots,M(n-1)\}$\footnote{
As usual, we assume that the input graph has non negative weight cycles. This can also be verified
initially using known algorithms such as \cite{YuZw05} in running times that are faster than the claimed running time of our algorithm.}. 
Using binary search one can locate the diameter which is the smallest $d$ such that
Threshold APSP reports all ordered pairs of vertices.
Observe that Threshold APSP is an interesting problem on its own. Especially interesting are the cases $d=0$
and $d=-1$ in the negative weight setting. The case $d=0$ can be used to find all pairs that have positive distance 
and the case $d=-1$ can be used to find all pairs that have negative distance. Using two consecutive values of $d$
one can also locate all pairs having any specific given distance.
Our main results are, therefore, algorithms for Threshold APSP.
\begin{theorem}
\label{t:main}
Let $G=(V,E)$ be directed graph with $n$
vertices and having integer edge weights in $\{-M,\ldots,0,\ldots,M\}$ and let $d$ be an integer.
Then, with high probability, the set of all ordered pairs of vertices having distance at most $d$ can be computed in
$\Ot(M^{\w/(\w+1)}n^{(\w^2+3)/(\w+1)})$ time. In particular, the diameter of $G$ can be computed in
$\Ot(M^{\w/(\w+1)}n^{(\w^2+3)/(\w+1)})$ time.
\end{theorem}
For $\w=2.376$ and for bounded edge weights, the running time of our algorithm is $O(n^{2.561})$.
It is, therefore, polynomially faster than the $O(n^{2.575})$ algorithm of Zwick. An even larger gap exists if $\w=2+o(1)$, as may turn out to be the case.
In this case, Zwick's algorithm, as well as an earlier algorithm of Alon, Galil, and Margalit \cite{AlGaMa97}, run in $\Ot(n^{2.5})$ time.
It is, thus, plausible that $\Omega(n^{2.5})$ is a barrier for APSP in directed graphs.
Our algorithm, on the other hand, computes the diameter in $\Ot(n^{7/3})$ time, assuming $\w=2+o(1)$.
Our result thus intensifies the plausibility of a true complexity gap between Diameter and APSP in weighted directed graphs.

For unbounded edge weights, our algorithm is also faster than Zwick's $\Ot(M^{1/(4-\w)}n^{2+1/(4-\w)})$ algorithm
(also if fast rectangular matrix multiplication is used in the latter) as long as $M < n^{3-\w}$.
But observe that for $M \ge n^{3-\w}$ both algorithms become cubic and it is preferable to use Chan's general APSP
algorithm in this case.

The main idea of our algorithm is as follows.
We start by computing values that capture the actual distance between all pairs that are ``far apart''
(pairs whose distance is realized only by a path that has relatively many edges). 
This is done using the well-known ``long shortest path'' technique.
The more difficult case is to capture distances connecting pairs that are not far apart.
For this purpose we use an appropriately computed {\em redundant partial distance matrix} that is a generalization of partial distance matrices introduced in \cite{YuZw05}.
A scaled version of this matrix enables us to obtain a good {\em additive} approximation for the distance between such pairs. If the approximated value is sufficiently smaller than the threshold $d$ then we already know that such pairs must be reported. For pairs whose approximation is close to $d$ we can use a truncated and shifted version of the redundant partial distance matrix in order to compute their distance precisely.

We now state our result for the positive integer weighted case.
\begin{theorem}
\label{t:main-2}
Let $G=(V,E)$ be directed graph with $n$
vertices and having integer edge weights in $\{1,\ldots,M\}$ and let $d$ be an integer.
Then, the set of all ordered pairs of vertices having distance at most $d$ can be computed in
$\Ot(Mn^\w)$ time. In particular, the diameter of $G$ can be computed in $\Ot(Mn^\w)$ time.
\end{theorem}

Similar to other shortest paths algorithms, both of our algorithms for Threshold APSP
can actually construct a {\em path data structure}. For two vertices $u,v$ reported as having distance at most $d$,
we can generate an actual path from $u$ to $v$ whose total weight is at most $d$ in time that is proportional to
the number of edges of the generated path. Observe, however, that for the purpose of computing the diameter,
we do not even need to construct such a data structure. Once we have computed the diameter value $d$, we can
use the results of Threshold APSP with the threshold $d-1$ in order to locate all pairs of vertices that realize the
diameter precisely. These are all pairs reported for the threshold $d$ and not reported for the threshold $d-1$.
Now, if we pick any such pair $u,v$, we can perform a Single Source Shortest Paths computation from $u$ in
$\Ot(Mn^\w)$ time (see, e.g. \cite{YuZw05,Sa05}) and obtain an actual path that realizes the diameter.

The rest of this paper is organized as follows. The next section contains some preliminary definitions, notations, and known results that are required for the proof
of the theorems. Section 3 describes redundant partial distance matrices that are an important ingredient in
the proof of Theorem \ref{t:main}.
Section 4 contains the proof of Theorem \ref{t:main} and Section 5 contains the proof of Theorem \ref{t:main-2}.
The final section contains some concluding remarks.

\section{Preliminaries}\label{s:prelim}

Let $T(\ell,m,n)$ be the minimal number of algebraic operations
needed to compute the product of an $\ell\times m$ matrix by an
$m\times n$ matrix.

\begin{definition}[Matrix multiplication exponents]
Let $\w(r,s,t)$ be the infimum of all the
exponents~$\w'$ for which $T(n^r,n^s,n^t)=O(n^{\w'})$. We
let $\w=\w(1,1,1)$ be the exponent of square matrix
multiplication.
\end{definition}

\begin{theorem}[Coppersmith and Winograd \cite{CoWi90}]
\label{t:CW}
$\w<2.376$.
\end{theorem}
In fact, Coppersmith \cite{Co97} proved that if $\alpha$ is the supremum over all constants
$r$ for which $\w(1,r,1)=2$ then $\alpha>0.294$.
The following lemmas are obtained by decomposing a given matrix
product into smaller products. (See, e.g., Huang and Pan \cite{HuPa98}.)

\begin{lemma}
\label{l:1}
$\w(1,r,1) \le \cases{ \hfill 2 \hfill & if $0\le r\le
\alpha$,\cr 2+\frac{\w-2}{1-\alpha}(r-\alpha) & if $\alpha\le
r\le 1$.\cr}$
\end{lemma}

\begin{lemma}
\label{l:2}
$\w(1,r,r)=\w(r,r,1)\le 1+(\w-1)r$, for $0\le r\le 1$.
\end{lemma}

\begin{definition}[Distance products]
Let $A$ be an $\ell\times m$ matrix, and let $B$
be a $m\times n$ matrix. Their $\ell\times n$ \emph{distance
product} $C=A\star B$ is defined as follows:
$C[i,j]=\min_{k=1}^m \{A[i,k]+B[k,j]\}$, for $1\le i\le \ell$ and
$1\le j\le n$.
\end{definition}

It is easy to see that if $W$ is an $n\times n$ matrix containing
the edge weights of an $n$-vertex graph, then $W^n$, the $n$-th
power of~$W$ with respect to distance products, is the {\em distance matrix} of the graph
(recall that we assume no negative weight cycles).
Namely, $W^n[i,j]=\delta(i,j)$ where $\delta(i,j)$ denotes the distance from $i$ to $j$.

As the fast algebraic matrix multiplication algorithms rely heavily
on the ability to perform \emph{subtractions}, they cannot be used
directly for the computation of distance products. Nevertheless, we
can get the following result, first stated by Alon et al.
\cite{AlGaMa97}, following a related idea of Yuval \cite{Yu76}.

\begin{lemma}
\label{l:dp} Let $A$ be an $n^r\times n^s$ matrix and let $B$ be an
$n^s\times n^t$ matrix, both with elements taken from
$\{-M,\ldots,0,\ldots,M\}\cup\{+\infty\}$. Then, the distance
product $A\star B$ can be computed in $\Ot(Mn^{\w(r,s,t)})$ time.
\end{lemma}

\begin{definition}[Truncation]
If $D$ is a matrix and $t$ is a positive integer, let
$\langle D \rangle_t$ be the matrix obtained from~$D$ by
replacing all the entries that are larger than~$t$ or smaller than $-t$ by $+\infty$.
In other words, $\langle D \rangle_t[i,j]=D[i,j]$, if $|D[i,j]|\le t$, and
$\langle D \rangle_t[i,j]=+\infty$, otherwise.
\end{definition}

Zwick \cite{Zw02} used several novel ideas combining truncated distance products, the notion of bridging sets,
and fast matrix multiplication, to obtain the presently fastest APSP algorithm in dense directed graphs
with integer edge weights.
\begin{theorem}[Zwick \cite{Zw02}]
\label{t:zwick}
Let $G$ be a directed graph with $n$ vertices with integer edge weights in $\{-M,\ldots,0,\ldots,M\}$.
Let $M=n^t$.
There is an algorithm that computes the distance matrix of $G$ in
$\Ot(n^{2+\mu(t)})$ time, where $\mu = \mu(t)$
satisfies the equation $\w(1,\mu,1)=1+2\mu-t$.
In particular, the algorithm runs in $O(n^{2.575})$ time if $~M$ is bounded.
\end{theorem}

\begin{definition}[Partial distance matrix]
A {\em partial distance matrix} of a graph $G$ is a matrix $P$ such that $P \star P$ is the distance matrix of $G$.
\end{definition}
Generalizing the above result of Zwick, it was shown by the author and Zwick how a {\em partial distance matrix} can be computed
much faster. Observe that once a partial distance matrix is computed, the distance between any given pair of vertices can be computed in $O(n)$ time.
\begin{theorem}[Yuster and Zwick \cite{YuZw05}]
\label{t:yz}
Let $G$ be a directed graph having $n$ vertices and integer edge weights in $\{-M,\ldots,0,\ldots,M\}$.
There exists an algorithm that computes a partial distance matrix of $G$ in
$\Ot(Mn^\w)$ time.
\end{theorem}

\section{Redundant partial distance matrices}

For two vertices $u,v$ of a graph, let $c(u,v)$ denote the smallest number of edges in a path that realizes $\delta(u,v)$.
 
\begin{definition}[$(\beta,\gamma)$-redundant partial distance matrix]
Let $G=(V,E)$ be a directed graph. For nonnegative parameters $\beta,\gamma$ with $0 \le \beta+\gamma \le 1$ we say that a matrix $P$
whose rows and columns are indexed by $V$ is a {\em $(\beta,\gamma)$-redundant partial distance matrix} ($(\beta,\gamma)$-RPDM) if
the following holds:
\begin{enumerate}
\item
For each pair $u,v \in V$ with $c(u,v) \le n^{1-\beta}$ there exists some $x \in V$ such that $P[u,x]+P[x,v] = \delta(u,v)$. Furthermore:
\item
There exists a path $p_{u,v}$ with $c(u,v)$ edges realizing $\delta(u,v)$ such that each segment of $p_{u,v}$ consisting of $\lceil n^{1-\beta-\gamma} \rceil$ edges contains a vertex $x$ such that $P[u,x]+P[x,v] = \delta(u,v)$. 
\end{enumerate}
\end{definition}

\newcommand{\PARTIAL}{\mbox{\bf redundant-partial-distance-matrix}}
\newcommand{\GETMIN}{\smash{\mathop{\longleftarrow}\limits^{\min}}}
\newcommand{\SAMPLE}{{\bf sample}}

\newcommand{\ALGORITHM}{{\tt algorithm}}
\newcommand{\FOR}{{\tt for}}
\newcommand{\ENDFOR}{{\tt endfor}}
\newcommand{\TO}{{\tt to}}
\newcommand{\DO}{{\tt do}}
\newcommand{\BEGIN}{{\tt begin}}
\newcommand{\IF}{{\tt if}}
\newcommand{\THEN}{{\tt then}}
\newcommand{\ELSE}{{\tt else}}
\newcommand{\END}{{\tt end}}
\newcommand{\ENDIF}{{\tt endif}}
\newcommand{\RETURN}{{\tt return}}

\newcommand{\ind}{\ell}

\begin{figure}[t]
\begin{center}\hspace*{-5pt}
\framebox{\hspace{0.4cm}\parbox{3.3in}{
\begin{tabbing}
\ALGORITHM$\;$ \PARTIAL$(W_{n\times n},\beta,\gamma)$ $\phantom{2^{2^{2^2}}}$ \\[5pt]
$V\gets \{1,2,\ldots,n\}$\\
$B\gets V$ ; $P\gets W$ \\[3pt]
\FOR\ \= $\ind\gets 1$ \TO\ $\lceil\log_{3/2}(n^{1-\beta-\gamma})\rceil$\\[3pt]
\> $s\gets (3/2)^\ell$\\[3pt]
\> $B\gets \SAMPLE(B,(9n\ln n)/s)$\\[3pt]
\> $P[V,B]\; \GETMIN\; \langle P[V,B]\rangle_{sM}\;\star\;\langle P[B,B]\rangle_{sM}$\\[3pt]
\> $P[B,V]\; \GETMIN\; \langle P[B,B]\rangle_{sM}\;\star\;\langle P[B,V]\rangle_{sM}$\\[3pt]
\ENDFOR\\[3pt]
\FOR\ \= $\ind\gets \lceil\log_{3/2}(n^{1-\beta-\gamma})\rceil+1$ \TO\ $\lceil\log_{3/2}(2n^{1-\beta})\rceil$\\[3pt]
\> $s\gets (3/2)^\ell$\\[3pt]
\> $P[V,B]\; \GETMIN\; \langle P[V,B]\rangle_{sM}\;\star\;\langle P[B,B]\rangle_{sM}$\\[3pt]
\> $P[B,V]\; \GETMIN\; \langle P[B,B]\rangle_{sM}\;\star\;\langle P[B,V]\rangle_{sM}$\\[3pt]
\ENDFOR\\[3pt]
\RETURN\ \= $P$
\end{tabbing}
}\hspace{0.4cm}}
\end{center}
\vspace*{-0.3cm} \caption{\label{f:rdd}{Computing a $(\beta,\gamma)$-redundant partial distance matrix}.}
\end{figure}

The algorithm given in Figure \ref{f:rdd} computes a $(\beta,\gamma)$-RPDM of a directed graph $G=(V,E)$ with integer edge weights in $\{-M,\ldots,0,\ldots,M\}$ whose adjacency matrix is
given as the input parameter $W$ (non-edges represented by $+\infty$ in $W$).
For subsets of vertices $X$ and $Y$, the notation $P[X,Y]$ appearing in the algorithm denotes the sub-matrix of $P$ consisting of the rows $X$ and columns $Y$.
For matrices $R$ and $S$ with the same dimensions, the notation $R\; \GETMIN\; S$ denotes the the assignment to $R$ of the matrix whose entry $[u,v]$ is the minimum of $R[u,v]$ and $S[u,v]$.
The algorithm in Figure \ref{f:rdd} is a modification of the algorithm from \cite{YuZw05} for computing partial distance matrices.
In fact, the first $\FOR$ loop (consisting of $\lceil\log_{3/2}(n^{1-\beta-\gamma})\rceil$ iterations) is identical to the algorithm in \cite{YuZw05}.
The second $\FOR$ loop differs from the first one in that $B$ remains constant and is no longer decreased by sampling.
We next prove the correctness and compute the running time of the algorithm $\PARTIAL$.

\begin{lemma}
\label{l:redundant-correct}
With high probability (at least $1-O(\log n/n)$), $\PARTIAL$ correctly computes a $(\beta,\gamma)$-RPDM.
\end{lemma}
\Proof
The first part of the proof is identical to the proof in \cite{YuZw05}. Let $B_\ell$ denote the random subset $B$ of the $\ell$-th iteration
(observe that when $\ell \le \lceil\log_{3/2}(n^{1-\beta-\gamma})\rceil$ we are in the first $\FOR$ loop and otherwise we are in the
second $\FOR$ loop).
We first note that Lemma 4.1 of \cite{YuZw05} remains intact: If $i \in B_\ell$ or $j \in B_\ell$, and there is a shortest
path from $i$ to $j$ in $G$ that uses at most $(3/2)^\ell$ edges, then after the $\ell$-th iteration, with
high probability (at least $1-O(\log n/n)$) we have $P[i,j] = \delta(i,j)$ (recall that the only difference between our algorithm
and the algorithm from \cite{YuZw05} is that we stop sampling from $B$ after the end of the first $\FOR$ loop; this is of course to our
advantage since the probability of hitting a path with a larger sample is larger).

Let $u$ and $v$ be two vertices with $c(u,v) \le n^{1-\beta}$, and let $p_{u,v}$ be a path with $c(u,v)$ edges from $u$ to $v$ realizing $\delta(u,v)$.
To establish the first part in the definition of a $(\beta,\gamma)$-RPDM, we only need to show that at the end of the algorithm there is,
with high probability, some $x \in B_\ell$ on $p_{u,v}$. This is identical to the proof of Lemma 4.2 in \cite{YuZw05}
(as shown there, such an $x$ already appears in $B_\ell$ where $(3/2)^\ell \ge c(u,v)$).
But in order to satisfy the second part in the definition of a $(\beta,\gamma)$-RPDM we need to prove something stronger:
we must show that, with high probability,
any segment of $n^{1-\beta-\gamma}$ edges of $p_{u,v}$ contains some $x \in B_\ell$ at the end of the algorithm.
As $B_\ell$ remains the same after the last iteration of the first $\FOR$ loop, we must prove that a random subset of size
$9n\ln n/n^{1-\beta-\gamma}$ vertices hits every such segment of $p_{u,v}$. Indeed, the probability that no vertex of
$B_\ell$ hits a specific segment is at most
$$
(1-9\ln n/n^{1-\beta-\gamma})^{n^{1-\beta-\gamma}} < n^{-9}\;.
$$
As there are less than $n$ segments to consider in $p_{u,v}$, and as there are less than $n^2$ pairs of vertices $u,v$ to consider, we have that
with high probability (larger than $1-O(\log n/n)$), the set $B_\ell$ at the end of the first $\FOR$ loop (and hence also at the end of the
algorithm; it is the same set) hits every segment of $n^{1-\beta-\gamma}$ edges of each $p_{u,v}$ with $c(u,v) \le n^{1-\beta}$.
\qed

\begin{lemma}
\label{l:redundant-runtime}
$\PARTIAL$ runs in $\Ot(Mn^\w+ Mn^{2+(\w-2)(\beta+\gamma)+\gamma})$ time.
\end{lemma}
\Proof
There are a logarithmic number of iterations, and the most time consuming operation in each iteration is the computation of a distance product.
By Lemma \ref{l:dp}, a distance product in the first $\FOR$ loop can be computed in
$\Ot(sM \times T(n,n/s,n/s))$ time. By Lemma \ref{l:2}, multiplying an $n \times n/s$ matrix with an $n/s \times n/s$ matrix requires
$O(s(n/s)^\w)$ operations. Hence, the time to perform a single distance product is $\Ot(Ms^2 (n/s)^\w)$.
Since $\w \ge 2$, this is never larger than $\Ot(Mn^\w)$ (this is the same argument that is used in \cite{YuZw05} to show that their
algorithm runs in $\Ot(Mn^\w)$ time). By Lemma \ref{l:dp}, a distance product in the second $\FOR$ loop can be computed in
$\Ot(sM n^{\w(1,\beta+\gamma,\beta+\gamma)})$. This is maximized in the last iteration where $s = \Theta(n^{1-\beta})$.
By Lemma \ref{l:2} this amounts to $\Ot(Mn^{2+(\w-2)(\beta+\gamma)+\gamma})$.
\qed

\section{Proof of Theorem \ref{t:main}}\label{s:main}

Let $G=(V,E)$ be a directed graph with $n$ vertices and with integer edge weights taken from
$\{-M,\ldots,0,\ldots,M\}$. Recall that $G$ is assumed to contain no negative weight cycles.
Let $d$ be any integer and let $D = \{(u,v)~:~ \delta(u,v) \le d\}$.
Our goal is to construct the set $D$. Observe that if $d < -nM$ then $D = \emptyset$ (otherwise there are negative
cycles). Also, if $d > nM$ then $D$ is precisely the set of pairs $(u,v)$ with $\delta(u,v) < \infty$ and
in this case $D$ can trivially be obtained from the transitive closure of the unweighted version of $G$.
Hence we will assume that $|d| \le nM$.   

\subsection{Retrieving distances that are realized by long paths}

The first part of our algorithm computes the distances between all pairs of vertices for which $c(u,v)$ is large.
More precisely, let $\beta$ be a chosen such that
\begin{equation}
\label{e:beta}
n^\beta = M^{\frac{w}{\w+1}}n^{\frac{(\w-1)^2}{\w+1}}
\end{equation}
and let $t=n^{1-\beta}$.
We compute, for {\em each} ordered pair of vertices $(u,v)$, and with very high probability,
a value $\delta_t(u,v)$ which satisfies $\delta_t(u,v) \ge \delta(u,v)$ and if $c(u,v) \ge t$ then
$\delta_t(u,v)=\delta(u,v)$.

\begin{lemma}
\label{l:large-c}
There is an algorithm that computes, with probability at least $1-O(1/n)$, values $\delta_t(u,v)$ for each ordered pair of vertices $(u,v)$ such that
$\delta_t(u,v) \ge \delta(u,v)$ and if $c(u,v) \ge t$ then $\delta_t(u,v) = \delta(u,v)$.
The algorithm runs in $\Ot(n^{2+\beta}+Mn^\w)$ time.
\end{lemma}
\Proof
Our algorithm uses the well-known ``long shortest path'' method to compute
the distances between pairs whose distance is realized only by paths with at least $t$ edges.
The idea behind this simple method is that a random large subset $X \subset V$
hits all of these ``long'' shortest paths.

More formally, let
$$
C = \{ (u,v) : c(u,v) \ge t \}.
$$
For each pair in $C$, fix a shortest path $p_{u,v}$ from $u$ to $v$.
Each such path contains at least $t+1$ vertices (including endpoints).
Let $X$ be a random subset of $8n\ln n/t$ vertices. For a pair $(u,v) \in C$, what is the probability that
no element of $X$ lies on $p_{u,v}$? Clearly, this probability is at most
$$
\left(1-\frac{c(u,v)+1}{n} \right)^{|X|} < \left(1-\frac{t}{n} \right)^{|X|} \le \left(1-\frac{t}{n} \right)^{8n\ln n/t} < \frac{1}{n^3}\;.
$$
As $|C| < n^2$ is follows by the union bound that with high probability (at least $1-O(1/n)$), $X$ intersects each $p_{u,v}$ for
$(u,v) \in C$.

So, assume that $X$ is a set of $8n\ln n/t$ vertices intersecting all $p_{u,v}$ for $(u,v) \in C$.
For each $x \in X$ our next goal is to compute, using a single-source (SSSP) algorithm, all the
distances $\delta(x,v)$ and all the distances $\delta(v,x)$ for all $v \in V$.

Unfortunately, $G$ has negative edge weights so performing $|X|$ applications of SSSP is too costly.
As observed by Johnson \cite{Jo77}, by an appropriate reweighing,
we can settle for just {\em one} application of SSSP and then reduce
the problem to SSSP in a graph with non-negative edge weights.
Johnson's reweighing consists of running a {\em single} application of SSSP from a new vertex, denoted by $r$, connected
with directed edges of weight $0$ from $r$ to each vertex of $V$.
An $\Ot(Mn^\w)$ time SSSP algorithm for directed graphs with integer weights in $\{-M,\ldots,0,\ldots,M\}$ was obtained in \cite{Sa05,YuZw05}.
It follows that the required reweighing of $G$ can be obtained in $\Ot(Mn^\w)$ time.
The reweighing consists of assigning vertex weights $h(v)$ for each $v \in V$
(these are the distances from $r$ to $v$ after applying SSSP from $r$).
The new weight, denoted by $w_+(u,v)$ is just $w(u,v)+h(u)-h(v) \ge 0$.
It now suffices to compute $SSSP$ from each vertex of $X$ in $G$ (and similarly in its edge-reversed version).
This, in turn, can be performed in $O(n^2)$ time for each vertex of $X$, using Dijkstra's algorithm.

Altogether, the running time required to obtain all of the distances $\delta(x,v)$ and $\delta(v,x)$ for all $x \in X$ and $v \in V$
is
$$
\Ot(n^2|X|+Mn^\w) = \Ot(n^3/t + Mn^\w) = \Ot(n^{2+\beta} + Mn^\w)\;.
$$
Next, for each ordered pair of vertices $(u,v)$, we compute
$$
\delta_t(u,v) = \min_{x \in X} \delta(u,x) + \delta(x,v)\;.
$$
Observe that the right hand side of the last equation is either infinity or a weight of {\em some} walk from $u$ to $v$
and thereby $\delta_t(u,v) \ge \delta(u,v)$. But for pairs $(u,v) \in C$ the property of $X$ guarantees that,
in fact, $\delta_t(u,v) = \delta(u,v)$, as required.
 
The running time to obtain all of the values $\delta_t(u,v)$ is $O(n^2|X|) \le \Ot(n^{2+\beta})$.
The overall running time of the algorithm is therefore $\Ot(n^{2+\beta} + Mn^\w)$, as claimed.
\qed

\subsection{Obtaining a good additive approximation}

The second part of our algorithm computes {\em approximate} distances between all pairs of vertices for which $c(u,v)$ is
relatively small. This process consists of a logarithmic number of steps, where each step guarantees to approximate
distances $\delta(u,v)$ for a specific range of $c(u,v)$.
More precisely, for $i=0,\ldots, \lfloor (1-\beta)\log n \rfloor$, let $t_i$ and $\beta_i$ be defined by
$$
t_i = n^{1-\beta_i} = \frac{n^{1-\beta}}{2^i}
$$
and observe that $\beta_0=\beta$ and $t_0=t$.
Step $i$ computes, for {\em each} ordered pair of vertices, and with very high probability,
a value $\delta_i^*(u,v)$ satisfying $\delta_i^*(u,v) \ge \delta(u,v)$ and
if $t_i/2 \le c(u,v) < t_i$ then $\delta_i^*(u,v) \le \delta(u,v)+2k_i$ for a suitably chosen
approximation parameter $k_i$. Observe that for any pair $(u,v)$ with $c(u,v) < t$, there exists {\em some} $i$ for which
$t_i/2 \le c(u,v) < t_i$.

Let $\gamma_i$ be defined by
\begin{equation}
\label{e:gamma}
n^{\gamma_i} = \left(n^{1-\beta_i}\right)^{\frac{\w-1}{\w}}
\end{equation}
and observe that $n^{\beta_i+\gamma_i} \le n$ so $\beta_i+\gamma_i \le 1$.
Since $\beta_0=\beta$ we also define $\gamma=\gamma_0$.

Let $P_i$ be a $(\beta_i,\gamma_i)$-RPDM computed by the algorithm in Section 3.
We use $P_i$ to obtain the claimed additive approximation.

\newcommand{\ADDITIVE}{\mbox{\bf additive-approximate}}
\begin{figure}[t]
\begin{center}\hspace*{-5pt}
\framebox{\hspace{0.4cm}\parbox{3.3in}{
\begin{tabbing}
\ALGORITHM$\;$ \ADDITIVE$(P_i,\gamma_i,k_i)$ $\phantom{2^{2^{2^2}}}$ \\[5pt]
$R_i\gets P_i/k_i$\\
$X\gets \SAMPLE(V,12n^{1-\gamma_i}\log n)$\\[3pt]
$Q_i\gets R_i[V,X]\;\star\;R_i[X,V]$\\[3pt]
\RETURN\ \= $k_iQ_i$
\end{tabbing}
}\hspace{0.4cm}}
\end{center}
\vspace*{-0.3cm} \caption{\label{f:additive}{Obtaining an additive approximation}.}
\end{figure}

Algorithm $\ADDITIVE$ in Figure \ref{f:additive} accepts $P_i$ and $\gamma_i$ as input, as well as an approximation parameter $k_i$.
It returns a matrix such that with high probability, its entry $[u,v]$ is close to $\delta(u,v)$ whenever  $t_i/2 \le c(u,v) < t_i$.

\begin{lemma}
\label{l:additive}
With probability $1-O(1/n)$, the matrix $k_iQ_i$ returned by $\ADDITIVE$ has the property that for
each ordered pair of vertices $(u,v)$ we have  $k_iQ_i[u,v] \ge \delta(u,v)$ and if $t_i/2 \le c(u,v) < t_i$ then
$k_iQ_i[u,v] \le \delta(u,v) + 2k_i$. Using $k_i=Mn^{1-\beta_i-\gamma_i}$ the running time of $\ADDITIVE$ is
$\Ot(n^{\gamma_i+\w(1,1-\gamma_i,1)})$.
\end{lemma}
\Proof
Consider an ordered pair of vertices $(u,v)$ for which $t_i/2 \le c(u,v) < t_i = n^{1-\beta_i}$.
By the definition of $P_i$, we have that there exists a path $p_{u,v}$ with $c(u,v)$ edges
realizing $\delta(u,v)$ where each segment of $\lceil n^{1-\beta_i-\gamma_i} \rceil$ edges of $p_{u,v}$ contains a vertex $x$
such that $P[u,x]+P[x,v] = \delta(u,v)$. In particular, there are at least
$$
\frac{c(u,v)}{\lceil n^{1-\beta_i-\gamma_i} \rceil} \ge \frac{t_i}{4n^{1-\beta_i-\gamma_i}} =  \frac{n^{\gamma_i}}{4}
$$
such vertices $x$.
This, in turn, implies that a random subset $X$ of $12n^{1-\gamma_i}\log n$ vertices contains, with high probability (at least
$1-O(1/n)$) at least one such $x$
for each pair $(u,v)$ with $t_i/2 \le c(u,v) < t_i$.
This means that, w.h.p., the entry $[u,v]$ of the product $P_i[V,X] \star P_i[X,V]$ contains the {\em precise} value of
$\delta(u,v)$ for such a pair $(u,v)$. Performing this distance product is, however, costly.
Instead, we divide each finite element in $P_i$ by the number $k_i$.
More precisely, the line $R_i\gets P_i/k_i$ denotes that $R_i[u,v] = \lceil P_i[u,v]/k_i \rceil$ for all elements of $P$.
We have
$$
P_i[u,x]+P_i[x,v] \le k_i(R_i[u,x] + R_i[x,v]) \le P_i[u,x]+P_i[x,v] + 2k_i
$$
and in particular, for all $(u,v)$ with $t_i/2 \le c(u,v) < t_i$,
$$
\delta(u,v) \le k_iQ_i[u,v] \le \delta(u,v)+2k_i\;. 
$$
Notice that for other pairs, the value $k_iQ_i[u,v]$ is {\em at least} $\delta(u,v)$ since the entries in $R_i$ are rounded up.

When applying $\ADDITIVE$ we will use $k_i=Mn^{1-\beta_i-\gamma_i}$.
Observe that the finite entries in $P_i$ have value $O(Mn^{1-\beta_i})$ and hence the finite entries in $R_i$ have value $O(n^{\gamma_i})$.
The running time of $\ADDITIVE$ is dominated by the distance product of an $n \times |X|$ sub-matrix of $R$ with an $|X| \times n$ sub-matrix
of $R$. Hence, by Lemma \ref{l:dp} and since $|X| =  \Ot(n^{1-\gamma_i})$, we have that $\ADDITIVE$ runs
in $\Ot(n^{\gamma_i+\w(1,1-\gamma_i,1)})$ time.
\qed

%Finally, by Lemma \ref{l:additive} we can take the largest entry of $kQ$ ranging over all $(u,v) \in C$ and obtain
%a value $d^*$  such that $d-2n^{1-\beta-\gamma} \le d^* \le d$. We therefore have:
%\begin{corollary}
%\label{c:additive}
%Given a $(\beta,\gamma)$-RPDM, there is an algorithm that, with high probability, computes for each $(u,v) \in C$ a value $\delta^*(u,v)$
%satisfying $\delta(u,v) - 2n^{1-\beta-\gamma} \le \delta^*(u,v) \le \delta(u,v)$. In particular, it computes a value $d^*$ such
%that $d-2n^{1-\beta-\gamma} \le d^* \le d$. The running time of the algorithm is
%$\Ot(n^{\gamma+\w(1,1-\gamma,1)})$.
%\end{corollary}

\subsection{Targeting pairs having distance at most $d$}

The final part of our algorithm correctly reports all pairs $(u,v)$ with $\delta(u,v) \le d$.
In the previous parts of our algorithm we have computed, for each pair $(u,v)$, values
$\delta_t(u,v)$ and $\delta_i^*(u,v)$ for $i=0,\ldots, \lfloor (1-\beta)\log n \rfloor$.
Let
$$
\delta^*(u,v) = \min \{\delta_t(u,v)~,~ \min_{i=0}^{\lfloor (1-r)\log n \rfloor} \delta_i^*(u,v)\}\;. 
$$
The following lemma is a consequence of these computed values.
\begin{lemma}
\label{l:good-approx}
For any pair $(u,v)$ we have $\delta(u,v) \le \delta^*(u,v) \le \delta(u,v)+ 2Mn^{1-\beta-\gamma}$. 
\end{lemma}
\Proof
Recall that all computed values are either infinite or upper bounds of weights of some walks from $u$ to $v$
and therefore $\delta(u,v) \le \delta^*(u,v)$. Now, if $c(u,v) \ge t$ then $\delta_t(u,v)=\delta(u,v)$.
Otherwise, there is some $i$ such that $t_i/2 \le c(u,v) < t_i$ in which case
$\delta_i^*(u,v) \le \delta(u,v)+2k_i$.
Since $k_i=Mn^{1-\beta_i-\gamma_i}$ the claim follows once we observe that $\beta+\gamma \le \beta_i+\gamma_i$.
Indeed this holds since by (\ref{e:gamma})
$$
n^{\beta_i+\gamma_i} = n^{\frac{\w-1+\beta_i}{\w}} \ge n^{\frac{\w-1+\beta}{\w}} =  n^{\beta+\gamma}\;.
$$
\qed

By Lemma \ref{l:good-approx}, any pair $(u,v)$ with $\delta^*(u,v) \le d$ is
reported as having $\delta(u,v) \le d$, as required.
Similarly, any pair with $\delta^*(u,v) > d+2Mn^{1-\beta-\gamma}$ is reported as having
$\delta(u,v) > d$, as required.
So we remain with the following set of pairs
$$
C^* = \{ (u,v) : d < \delta^*(u,v) \le d+2Mn^{1-\beta-\gamma} \}\;
$$
where we must determine, for each $(u,v) \in C^*$, whether or not $\delta(u,v) \le d$.
In fact, we will determine $\delta(u,v)$ precisely for all $(u,v) \in C^*$.

Let $C_i$ denote the set of pairs in $C^*$ for which $t_i/2 \le c(u,v) < t_i$.
Observe that we {\em do not} know $C_i$ and, furthermore, there may be pairs $(u,v) \in C^*$ that are in no
$C_i$ as it may be that $\delta^*(u,v)=\delta_t(u,v)$. 
Consider again the $(\beta_i,\gamma_i)$-RPDM matrix $P_i$.
By the definition of $P_i$, for each $(u,v) \in C_i$, there is a shortest path $p_{u,v}$ from $u$ to $v$ with $c(u,v)$ edges
such that every segment of $\lceil n^{1-\beta_i-\gamma_i} \rceil$ edges of
$p_{u,v}$ contains a vertex $x$ such that $P_i[u,x]+P_i[x,v]=\delta(u,v)$. In particular, there is such an $x$ where
$$
P_i[u,x], P_i[x,v] \in [\frac{\delta(u,v)}{2}-Mn^{1-\beta_i-\gamma_i} \;,\; \frac{\delta(u,v)}{2}+Mn^{1-\beta_i-\gamma_i}]\;.
$$
But for pairs $(u,v) \in C^*$ we have, in particular, that
$$
P_i[u,x], P_i[x,v] \in [\frac{d}{2}-2Mn^{1-\beta-\gamma} \;,\; \frac{d}{2}+2Mn^{1-\beta-\gamma}]\;.
$$

Let $S_i$ be the matrix obtained from $P_i$ by replacing each entry not in $[\frac{d}{2}-2Mn^{1-\beta-\gamma} \;,\; \frac{d}{2}+2Mn^{1-\beta-\gamma}]$
with $+\infty$, and by decreasing each remaining entry by $\lfloor \frac{d}{2}-2Mn^{1-\beta-\gamma} \rfloor$.
The distance product $R_i = S_i \star S_i$ has, therefore, the property that for each $(u,v) \in C_i$,
$$
R_i[u,v] + 2\lfloor \frac{d}{2}-2Mn^{1-\beta-\gamma} \rfloor = \delta(u,v)\;.
$$
\newcommand{\TARGET}{\mbox{\bf target-distances}}
\begin{figure}[t]
\begin{center}\hspace*{-5pt}
\framebox{\hspace{0.4cm}\parbox{3.3in}{
\begin{tabbing}
\ALGORITHM$\;$ \TARGET$(P_i,\beta,\gamma,d)$ $\phantom{2^{2^{2^2}}}$ \\[5pt]
\FOR\ \= $(u,v) \in V \times V$\\[3pt]
\> \IF\ \= $P_i[u,v] \in [\frac{d}{2}-2Mn^{1-\beta-\gamma} \;,\; \frac{d}{2}+2Mn^{1-\beta-\gamma}]$\\[3pt]
\> \> $S_i[u,v]\gets P_i[u,v]-\lfloor \frac{d}{2}-2Mn^{1-\beta-\gamma} \rfloor$\\[3pt]
\> \ELSE\\[3pt]
\> \> $S_i[u,v]\gets +\infty$\\[3pt]
$R_i \gets S_i\;\star\;S_i$\\[3pt]
\RETURN\ \= $R_i + 2\lfloor \frac{d}{2}-2Mn^{1-\beta-\gamma} \rfloor$
\end{tabbing}
}\hspace{0.4cm}}
\end{center}
\vspace*{-0.3cm} \caption{\label{f:target}{Computing distances for pairs in $C_i$}.}
\end{figure}
This procedure is summarized in Algorithm  $\TARGET$ given in Figure \ref{f:target}.
As each entry of $S_i$ has value $O(Mn^{1-\beta-\gamma})$ the distance product $S_i \star S_i$ takes $\Ot(Mn^{1-\beta-\gamma+\w})$ time.
The following lemma proves the correctness of our algorithm.
\begin{lemma}
\label{l:target}
For each $(u,v) \in C^*$ we have
$$
\delta(u,v) = \min\{ \delta_t(u,v) ~,~  \min_{i=0}^{\lfloor (1-r)\log n \rfloor} R_i[u,v] + 2\lfloor \frac{d}{2}-2Mn^{1-\beta-\gamma} \rfloor\}\;.
$$
Furthermore, all the values $\delta(u,v)$ for $(u,v) \in C^*$ are computed in $\Ot(Mn^{1-\beta-\gamma+\w})$ time.
\end{lemma}
\Proof
Let $(u,v) \in C^*$. If $c(u,v) \ge t$ then we already have $\delta_t(u,v) = \delta(u,v)$ as required.
Otherwise, for some $i$, $t_i/2 \le c(u,v) < t_i$ in which case we have shown that algorithm $\TARGET$ applied to $P_i$ gives
$$
R_i[u,v] + 2\lfloor \frac{d}{2}-2Mn^{1-\beta-\gamma} \rfloor = \delta(u,v)\;.
$$
Algorithm $\TARGET$ is applied $O(\log n)$ times, once for each $P_i$. As a single application of $\TARGET$ runs in $\Ot(Mn^{1-\beta-\gamma+\w})$ time, the claim regarding the running time follows.
\qed

\subsection{Running time}

It remains to show that all ingredients of our algorithm run in
$\Ot(M^{\w/(\w+1)}n^{(\w^2+3)/(\w+1)})$ time.

The first part of our algorithm, given as Lemma \ref{l:large-c}, runs in $\Ot(n^{2+\beta}+Mn^\w)$ time.
Recall that we can assume that $M \le n^{3-\w}$ (as otherwise the usual cubic algorithms are faster, as mentioned in
the introduction) so by (\ref{e:beta}),
$Mn^\omega = O(n^{2+\beta})$ and also
$$
n^{2+\beta} = M^{\frac{\w}{\w+1}}n^{2+\frac{(\w-1)^2}{\w+1}} = M^{\frac{\w}{\w+1}}n^\frac{\w^2+3}{\w+1}\;.
$$
Hence this part of the algorithm satisfies the claimed running time assertion.

Computing a $(\beta_i,\gamma_i)$-RPDM (and recall that we perform this $O(\log n)$ times) takes $\Ot(Mn^\w+ Mn^{2+(\w-2)(\beta_i+\gamma_i)+\gamma_i})$ time, as shown by Lemma \ref{l:redundant-runtime}. Using (\ref{e:gamma}) it is straightforward to verify that
$$
Mn^{2+(\w-2)(\beta_i+\gamma_i)+\gamma_i} = Mn^\w n^{(1-\beta_i)/\w}\;.
$$
As $\beta_i \ge \beta_0 = \beta$ the worst case occurs when $i=0$. Plugging in the value of $n^{\beta}$ from
(\ref{e:beta}) we obtain that
$$
Mn^{2+(\w-2)(\beta+\gamma)+\gamma} = M^{\frac{\w}{\w+1}}n^\frac{\w^2+3}{\w+1}\;.
$$
Hence this does not exceed the claimed running time assertion.

The next part is algorithm $\ADDITIVE$ whose running time is $\Ot(n^{\gamma_i+\w(1,1-\gamma_i,1)})$ as stated in Lemma \ref{l:additive}.
We can use Lemma \ref{l:1}, but even if we use the naive bound $\w(1,1-\gamma_i,1) \le 2+(\w-2)(1-\gamma_i)$ we can see that this part is not a bottleneck of the algorithm. Indeed, since $\gamma_i \le \gamma_0 = \gamma$ we have
$$
n^{\gamma_i+\w(1,1-\gamma_i,1)} \le n^{\gamma_i + 2+(\w-2)(1-\gamma_i)} = n^{(3-\w)\gamma_i+\w} 
\le n^{(3-\w)\gamma+\w} = n^{(3-\w)\frac{\w-1}{\w}(1-\beta)+\w} =
$$
$$
n^{(3-\w)\frac{\w-1}{\w}+\w}\left( M^{\frac{w}{\w+1}}n^{\frac{(\w-1)^2}{\w+1}} \right)^{-(3-\w)\frac{\w-1}{\w}}=
$$
$$
n^{\frac{\w^3 - 6\w^2+16\w-9}{\w+1}}M^{-\frac{(3-\w)(\w-1)}{\w+1}} \le n^{\frac{\w^2+3}{\w+1}}
$$
where the last inequality is valid for all $\w \le 3$ (and is a strict inequality for $2 < \w < 3$).

The final part is algorithm $\TARGET$ given as Lemma \ref{l:target}. It runs in $\Ot(Mn^{1-\beta-\gamma+\w})$ time.
Plugging in the values from (\ref{e:beta}) and (\ref{e:gamma}) we obtain
$$
Mn^{1-\beta-\gamma+\w} = M^{\w/(\w+1)}n^{(\w^2+3)/(\w+1)}\;.
$$
We have established the claimed running time of our algorithm, thereby concluding the proof of Theorem \ref{t:main}. \qed

\section{Proof of Theorem \ref{t:main-2}}\label{s:main-2}

For a nonnegative integer $k$ and for a positive integer $M$ we define a set of nonnegative integers $F(k,M)$ recursively as follows.
$$
F(k,M) =
\cases{
\hfill \{0,\ldots,k\} \hfill & if $k \le M+1$,\cr
\cr
\{k\} \bigcup_{i=\lfloor (k-M)/2 \rfloor}^{\lceil (k+M)/2 \rceil}F(i,M) & if $k > M+1$.\cr
}
$$
For example, $F(100,4)=\{100,52,\ldots,48,28,\ldots,22,16,\ldots,0\}$.
Observe that $F(k,M)$ consists of $O(\log(k+M))$ intervals of consecutive integers. 
\begin{lemma}
\label{l:bound}
$|F(k,M)| = O(M \log k)$.
\end{lemma}
\Proof
Assume that $k > 3M$ otherwise the claim clearly holds.
The $j$'th level of the recursion defining $F(k,M)$ consists of unions of sets $F(i,M)$
where $i$ is contained in the interval $[k/2^j-M-1,k/2^j+M+1]$.
As there are $O(log(M+k))$ levels of recursion, the claim follows.
\qed

For a directed graph $G=(V,E)$ with edge weights in $\{1,\ldots,M\}$ let $A_k$ denote the
Boolean matrix whose rows and columns are indexed by $V$, and $A_k[u,v]=1$ if and only if
$\delta(u,v) \le k$. Also define $A_0$ to be the identity matrix.
For a given threshold value $d$ our goal is to compute $A_d$.

Our algorithm will, in fact, compute all the matrices $A_i$ for $i \in F(d,M)$.
In particular, as $d \in F(d,M)$, we will eventually obtain the required $A_d$.

For $i \in F(d,M)$ we say that $i$ is {\em primal} if $i \le M+1$.
We say that $i$ belongs to {\em level} $j$ if $F(i,M)$ appears
in the $j$'th level of the recursion (observe that $i$ may belong to more than one level).
We denote by $L(d,M,j)$ the interval of consecutive integers forming the $j$'th level.
As observed in the proof of Lemma \ref{l:bound},
$L(d,M,j) \subset [d/2^j-M-1,d/2^j+M+1]$ so in particular $|L(d,M,j)| \le 2M+3$.
For example, consider again $F(100,4)$. Then $100$ is at level $0$.
Level $1$ consists of $\{48,\ldots,52\}$.
Level $2$ consists of $\{22,\ldots,28\}$.
Level $3$ consists of $\{9,\ldots,16\}$.
Level $4$ consists of $\{2,\ldots,10\}$.
Level $5$ consists of $\{1,\ldots,7\}$ (recursion continues only for $i > M+1$ so for $i \ge 6$ in this example).
Level $6$ consists of $\{1,\ldots,6\}$.
Finally, level $7$ consists of $\{1,\ldots,5\}$ and is the last level as it consists only of primal values. 

The following lemma essentially proves Theorem \ref{t:main-2}.
\begin{lemma}
\label{l:main-2}
Given all the matrices $A_i$ for $i \in L(d,M,j+1) \cup \{0,\ldots, M+1\}$
the set of all matrices $A_k$ for $k \in L(d,M,j)$ can be computed in $\Ot(Mn^\w)$ time. 
\end{lemma}
\Proof
Consider some matrix $A_k$ for $k \in L(d,M,j)$.
If $k \le M+1$ there is nothing to prove so assume that $k > M+1$.
Suppose that $u$ and $v$ are vertices such that $\delta(u,v) \le k$, and consider a shortest path from $u$ to $v$.
Let $w$ be the first vertex on this path for which $\delta(u,w) \ge \lfloor (k-M)/2 \rfloor$.
If there is no such vertex, then already $\delta(u,v) < \lfloor (k-M)/2 \rfloor$.
Otherwise, since each weight is in $\{1,\ldots,M\}$, then $\delta(u,w) \le  \lfloor (k-M)/2 \rfloor+M-1$.
Furthermore, $\delta(w,v) \le k - \lfloor (k-M)/2 \rfloor$. In any case
$A_k$ is just the Boolean {\bf or} of the following Boolean Matrix products:
\begin{equation}
\label{e:bool}
A_k = \vee_{i=\lfloor (k-M)/2 \rfloor}^{\lceil (k+M)/2 \rceil} A_iA_{k-i}\;.
\end{equation}
By the definition of $F(d,M)$, if $k \in L(d,M,j)$ then all the indices $i$ and $k-i$ in (\ref{e:bool})
belong to $L(d,M,j+1)$. Since Boolean matrix multiplication can be performed in $O(n^\omega)$ time,
(\ref{e:bool}) shows that we can compute $A_k$ in $O(Mn^\omega)$ time.

The only problem that remains is that we do not want to compute a single $A_k$.
We want all $A_k$ for all $k \in L(d,M,j)$ and as $|L(d,M,j)|=O(M)$ this takes $O(M^2n^\w)$ if
we compute each $A_k$ separately.  

To overcome this problem we use the following matrix convolution idea.
We construct a single matrix $B$ that encodes all the matrices $A_i$ for $i \in L(d,M,j+1)$ at once.
Set $s=|L(d,M,j+1)|$ and let $t$ be the smallest index in $|L(d,M,j+1)|$.
Recalling that $L(d,M,j+1)$ is an interval of consecutive integers we have
$L(d,M,j+1)=\{t,t+1,\ldots,t+s-1\}$.
Let $B[u,v]$ be the following polynomial of degree at most $s-1$.
$$
B[u,v] = \sum_{q=0}^{s-1}A_{q+t}[u,v]x^q\;.
$$
Thus, the coefficient of $x^q$ encodes the matrix $A_{q+t}$.

Now consider $C=B^2$ (product performed over the ring of polynomials in a single variable).
Each element of $C$ is therefore a polynomial of degree at most $2s-2$. Consider
the coefficient of $x^{k-2t}$ in $C[u,v]$. The only way it can be non-zero is if
for some $i$, $A_iA_{k-i}$ is nonzero in entry $[u,v]$. Hence, we obtain $A_k$ just
by examining the coefficients of $x^{k-2t}$ in the entries of $C$.  

The only thing that remains is to consider the complexity of computing $C$.
Two matrices whose entries are polynomials of degree at most $s-1$ and whose coefficients are
bounded integers (in our case the coefficients are either $0$ and $1$) can be multiplied
in $\Ot(sn^\w)$ time. The standard trick is to replace the variable $x$ with a large number (say $n+1$ if the dimension
of the matrices is $n$) so that no carry is introduced when reading the product entries as digits in base $n+1$,
and thereby constructing the polynomials in the entries of the product.
However, observe that replacing the variable $x$ with $n+1$ causes the entries to become as large as
$O(n^s)$ and hence consist of $O(s \log n)$ bits. Thus, each matrix operation incurs an $\Ot(s)$-factor
that cannot be ignored.

In our case we have $s=O(M)$ so we conclude that all the matrices $A_k$ for $k \in L(d,M,j)$ can be
computed in $\Ot(Mn^\w)$ time.
\qed

The proof of Theorem \ref{t:main-2} now follows from Lemma \ref{l:main-2} by recalling that the number of
levels is $O(\log(d+M))$.
The only thing that remains is to compute the matrices $A_1$, \ldots $A_{M+1}$ that correspond to
the primal indices. This, however, is relatively easy to do using distance products.
In fact, the following lemma is a consequence of the result of Zwick from \cite{Zw02}.

\begin{lemma}
\label{l:apspM}
Let $G$ be a graph with $n$ vertices and integer edge weights in $\{1,\ldots,M\}$.
There is an $\Ot(Mn^\w)$ time algorithm that computes $\delta(u,v)$ for all pairs
$(u,v)$ which satisfy $\delta(u,v) \le M+1$.
\end{lemma}
\Proof
Any pair that has $\delta(u,v) \le M+1$ has at most $M+1$ edges on any shortest path from $u$ to $v$.
The result of \cite{Zw02}, specifically, algorithm {\bf rand-short-path}, its complexity analysis, and Lemma 4.2
therein, show that the exact distances for such pairs is computed in $\Ot(Mn^\w)$ time.
We note that during the computations of the distance products in each iteration of {\bf rand-short-path} we never need to consider matrix entries with values
exceeding $M+1$. As the bridging sets of each iteration are of size $\Ot(n/s)$ (see Lemma 4.2 in \cite{Zw02}),
the time to compute the distance product of the rectangular matrices, even without resorting to fast rectangular matrix multiplications, is $\Ot(M(n/s)^\w s^2)$. As in our case we
always have $s=O(M)$, the result follows.
\qed

Observe that once we have $\delta(u,v)$ for all pairs $(u,v)$ with $\delta(u,v) \le M+1$ then we immediately
have the matrices $A_1,\ldots,A_{M+1}$, as required.
This concludes the proof that all $A_k$ for $k \in F(d,M)$ are computed in $\Ot(Mn^\w)$ time, and hence
the proof of Theorem \ref{t:main-2}. 

\section{Concluding remarks}

We presented an algorithm that computes the diameter (and Threshold APSP) of an integer weighted directed graph
polynomially faster than any presently known APSP algorithm.
The algorithm is randomized and returns,
with high probability,  the {\em precise} diameter, as well as all pairs of vertices realizing it.

Obtaining a truly subcubic algorithm for computing the diameter (moreover Threshold APSP) of real-weighted graphs remains an open problem. The prospects in this case,
however, seem gloomier. It is likely that Diameter and APSP are equivalent under sub-cubic reductions.
A recent result of Vassilevska Williams and Williams \cite{VaWi10} asserts that the existence of a truly subcubic algorithm for real-weighted APSP
is equivalent under subcubic reductions to the existence of truly subcubic algorithms for a list of problems that seem ``lighter'' than APSP.

\section*{Acknowledgments}
I thank Uri Zwick for some useful discussions. Special thanks to Andreas Bj\"orklund for fruitful discussions and
insightful comments.

\end{document}